\documentclass[conference]{IEEEtran}
\IEEEoverridecommandlockouts
\usepackage{cite}
\usepackage{amsmath,amssymb,amsfonts}
\usepackage{algorithmic}
\usepackage{graphicx}
\usepackage{textcomp}
\usepackage{xcolor}
\usepackage{array}

\usepackage{algorithm}
\usepackage{algorithmic}
\def\BibTeX{{\rm B\kern-.05em{\sc i\kern-.025em b}\kern-.08em
    T\kern-.1667em\lower.7ex\hbox{E}\kern-.125emX}}
\begin{document}

\title{Explainable AI in Request-for-Quote}

\author{\IEEEauthorblockN{1\textsuperscript{st} Qiqin Zhou}
\IEEEauthorblockA{
\textit{Operation Research and Information Engineering (Cornell University)}\\
Ithaca, New York \\
qz247@cornell.edu}

}

\maketitle

\begin{abstract}
In the contemporary financial landscape, accurately predicting the probability of filling a Request-For-Quote (RFQ) is crucial for improving market efficiency for less liquid asset classes. This paper explores the application of explainable AI (XAI) models to forecast the likelihood of RFQ fulfillment. By leveraging advanced algorithms including Logistic Regression, Random Forest, XGBoost and Bayesian Neural Tree, we are able to improve the accuracy of RFQ fill rate predictions and generate the most efficient quote price for market makers. XAI serves as a robust and transparent tool for market participants to navigate the complexities of RFQs with greater precision.

\end{abstract}


\section{Introduction}
Request-For-Quote (RFQ) is a form of trading where market participants can obtain electronic quotes on any asset class. RFQ allows for more efficient price discovery for less liquid asset classes such as To-be-Announcement (TBA) with a specific coupon. TBA is a forward contract on mortgage-backed securities that is widely used in the housing finance industry to protect against interest rate fluctuations during the mortgage lending process. Providing pricing for broader range of TBA bonds through RFQ is particularly advantageous for mortgage lenders, as traditional market makers typically only show the price of the most liquid coupons. The pricing from RFQ enhances the transparency in the TBA market and it in turn provides crucial data for generating precise valuation for borrowers in the primary mortgage origination market, ultimately offering mortgage borrowers better pricing and financing options in a tough market.

The objective of the paper is to create an algorithm that allows the market-maker to produce a right price of TBA to fill an RFQ while maximizing market-maker's utility and minimizing their inventory risk. The algorithm is naturally decomposed into two components: (a) evaluate the probability of filling an RFQ given a price and (b) determine the final price market-maker should set for the RFQ so as to maximize their utility.

In Section \ref{Data Generation Process}, we first proposed a simulation algorithm to produce hypothetical RFQ data that are otherwise not available to the public because of the regulatory constraints that limit the sharing and distribution of real RFQ data. Based on the synthesized dataset, we perform data analysis and feature engineering to disclose the pattern of RFQ fill rate in Section \ref{Data Processing and Feature Engineering}.

We address the problem of (a) in Section \ref{Model} by experimenting different machine learning and Explainable Artificial Intelligence (XAI) models to predict the probability of filling an RFQ. By leveraging a set of explainability techniques in machine learning and deep learning, we are able to enhance the transparency and reliability of predicting probability of filling the RFQ. It helps explain how each model input contributes to the final outcome. We compared various XAI models including decision trees, random forests and gradient boosting. We established a base case of traditional logistic regression models constrained by LASSO regularization. Our findings indicate that XAI methods can achieve comparable predictive performance while maintaining explainability, thus facilitating better decision-making in financial contexts. XAI has strong practical implications in helping financial institutions to establish a trading and risk management system compliant with regulatory requirements. It is also pivotal to improve the responsible use of AI in financial activities. 

To address problem (b) in Section \ref{price prediction}, we proposed a market-maker utility function that involves both the probability and profit margin of filling an RFQ. To predict the profit margin, we constructed a parsimonious model to predict the next middle price (mid-price) given the current information. Note that we do not intend to put emphasis on this price prediction model considering the lack of real-world intraday market data and the widely-acknowledged complexity of predicting the high noise-to-signal intraday price.

\section{Literature Review}
\label{Literature}
\subsection{RFQ}
The Request-For-Quote (RFQ) process allows market-makers to respond to trade requests from a set of counterparties. A counterparty initiates an RFQ detailing information on the trade request, which will typically include the instrument identifier, the size of the transaction, the side (buy or sell) and etc. The RFQ is sent to a selection of market-makers (where the counterparty will choose the optimal list based on past experience, hit-ratios). Upon receiving the RFQ, a market-maker will quote back a specific price at which the are willing to execute the trade (yet each market-maker is not aware of the prices from their competitor market-makers). When the counterparty is ready to execute, the market-maker with the best price will win the trade. Of note, the marketmaker will then have the other side of the trade and will need to manage the risk and hedge risk appropriately.

\subsection{Explainable Artificial Intelligence}
Predictive modeling in finance has evolved significantly with the advent of machine learning and Artificial Intelligence \cite{murphy2012machine, bishop2006pattern}. Traditional predictive models often prioritize accuracy over explainability, leading to black-box solutions that are difficult to understand and trust \cite{ribeiro2016should}. Explainable AI (XAI) addresses this issue by providing transparency in model predictions, which is essential for compliance with regulatory standards and for gaining the trust of market participants \cite{adadi2018peeking}. Recent advancements in XAI have made it possible to balance predictive power with explainability \cite{lundberg2017unified}. From the existing literature, we found there are two major categories of XAI: post-hoc explainability and intrinsic explainability.

Post-hoc explainability refers to model-agnostic techniques we can utilize to improve explainability. These include Local explainable Model-agnostic Explanations (LIME) \cite{ribeiro2016should}, SHAP (Shapley Additive explanations) values \cite{lundberg2017unified}, Partial Dependence Plots (PDPs) \cite{friedman2001greedy}, and Feature Permutation Importance \cite{breiman2001random}. However, they suffer from several shortcomings. The first problem is the approximation errors as many of these explainability techniques rely on approximations or simplifications of the model \cite{molnar2020explainable}. The next problem is scalability as applying these techniques to large datasets or complex models is extremely computationally challenging \cite{molnar2020explainable}. There are also limitations specific to each technique. For Feature Permutation Importance, the measurement of feature importance, such as Gini importance in Random Forests, can be biased towards features with more categories or higher cardinality \cite{strobl2007bias}. PDPs assume that features are independent of each other, which is often not true in real-world data, leading to misleading interpretations \cite{goldstein2015peeking}. The results of LIME are sensitive to how the local neighborhood around the prediction is defined, making LIME explanations unstable and providing different explanations for similar predictions if the local neighborhood changes slightly \cite{ribeiro2016should}.

Intrinsically explainable machine learning models are those whose structure and functioning are inherently understandable and explainable. Intrinsic model explainability is arguably the most desirable option as it addresses
all of the objectives to justify, control, discover, and improve. These models are typically simpler, and their predictions can be easily traced back to the input features. Logistic regression is one such explainable model that can be used for binary and multiclass classification \cite{hosmer2013applied}. The coefficients in a logistic model indicate the influence of each feature on the model outcome. Decision Tree or general tree models are also intrinsically explainable, as their tree nodes represent feature inputs, branches represent the decision rules, and leaves represent outcomes \cite{breiman1984classification}. The tree structure allows for a straightforward tracing of the decision-making process from the root to the leaf. In addition, Naive Bayes is a probabilistic classifier based on Bayes' theorem with a strong assumption of feature independence. Its decisions can be understood by examining the contributions of each feature to the posterior probabilities \cite{mccallum1998comparison}.

The above-mentioned models are all computationally efficient and explainable but often require careful feature engineering procedure to achieve a good performance as they are not as good at capturing complex, non-linear relationships in the data as more complex models can do. Additionally, unsupervised machine learning algorithms are also utilized as explainable model in finance such as using K-means to explicitly segment a covariance-tilted portfolio into sub-portfolios to maintain the stability of optimization \cite{zhou2024portfolio}.

To balance the trade-off between model performance and model explainability, we propose a new model framework: Bayesian Neural Tree (BNT). It is a hybrid model that integrates the explainability of decision trees with the powerful predictive capabilities of neural networks, framed within a Bayesian context to manage uncertainty and provide probabilistic predictions. The model leverages a tree structure where each node represents a neural network, allowing for complex decision boundaries while maintaining a hierarchical, explainable form. By incorporating Bayesian inference, BNTs can quantify uncertainty in predictions, making them particularly useful in applications requiring robust decision-making under uncertainty \cite{murphy2012machine}. This approach has shown promise in various fields, including finance and healthcare, where understanding the confidence in predictions is as crucial as the predictions themselves \cite{ghahramani2015probabilistic}.

\section{Data Generation Process}
\label{Data Generation Process}
Request-for-quote (RFQ) data is often not publicly available because it pertains to proprietary Information, data privacy and sensitivity. Therefore, we developed a simulation algorithm to synthetically construct a series of RFQ data for TBAs. In total we generated 10005 records where the last 5 records represent real-world RFQs we want to quote.

The variables we generated include: the timestamp of the initiation of the RFQ (\textit{Time}), the unique identified the the bond (Bond), the direction of the RFQ: offer or bid (\textit{Side}), bond notional (\textit{Notional}), the id of target counterparty (\textit{CounterParty}), the mid-price which is the average of level-1 bid and ask price (\textit{MidPrice}), the quoted price of RFQ (\textit{QuotedPrice}), number of competitors at the time (\textit{Competition}), the status of the RFQ: missed or done (\textit{Status}) and the mid price at in the following minute (\textit{NextMidPrice}). 

\textit{Time} is a continuous 5-digit integer that mimics the format of timestamp sent from the exchange. \textit{Notional} is an integer generated in $\log10$ scale from 3 to 7. \textit{Side} is binary variable and we assume it follows a  a Bernoulli distribution with probability \( p \) = 0.5. We assume there are in all 4 participants in RFQ so \textit{CounterParty} is generated uniformly from \([0,3]\). \textit{Competition} is generated uniformly from \([1,4]\) since there are at most 4 participants in the RFQ process.

\textit{Status} is the target of our prediction and is also a binary variable. We assume it follows a Bernoulli distribution that is independent of \textit{Side}. We used the following steps to generate it:
Let \( \mathbf{X} \) be defined as
\[
\mathbf{X} = 
\begin{pmatrix}
(1 + 0.3N) \cos(2\pi U) \\
(1 + 0.3N) \sin(2\pi U)
\end{pmatrix}
\]
where \( U \sim \text{Uniform}(0, 1) \) and \( N \sim \text{Normal}(0, 1) \) are independent random variables. The outcome \( Y \) follows a Bernoulli distribution with probability \( p \). This probability is a function of the Euclidean distance between \( X \) and \( S_1 \), namely
\[ 
p = \frac{1}{1 + e^{-10 \|X\|^2}} 
\]

The process for creating prices (\textit{QuotedPrice}, \textit{MidPrice},\textit{NextMidPrice}) is as follow: 

We assume the initial prices for the bid (\(B_0\)), ask (\(A_0\)), and mid-price (\(M_0\)) are defined as follows:
\[
B_0 = P_0 - \frac{S_0}{2}
\]
\[
A_0 = P_0 + \frac{S_0}{2}
\]
\[
M_0 = \frac{B_0 + A_0}{2}
\]
where \(P_0\) is the initial mid-price $124.24\$$ and \(S_0\) is the initial spread $0.1\$$.

The spread at each time step \( t \) then can be updated using a Geometric Brownian Motion model:
\[
S_{t+1} = S_t \exp\left(\mu \Delta t + \sigma_S \epsilon_t \sqrt{\Delta t}\right)
\]
where \(\mu\) is the drift coefficient, \(\sigma_S\) is the volatility coefficient, and \(\epsilon_t \sim \mathcal{N}(0,1)\) is a random shock drawn from a standard normal distribution.

At each time step \(t\), the bid and ask prices are updated using stochastic processes:
\[
B_{t+1} = M_t - \frac{S_{t+1}}{2} + \epsilon_B
\]
\[
A_{t+1} = M_t + \frac{S_{t+1}}{2} + \epsilon_A
\]
where \(\epsilon_B \sim \mathcal{N}(0, \sigma_B^2)\) and \(\epsilon_A \sim \mathcal{N}(0, \sigma_A^2)\) are random shocks drawn from normal distributions with mean zero and variances \(\sigma_B^2\) and \(\sigma_A^2\), respectively.

The mid-price at each time step \(t\), i.e \textit{MidPrice} is then given by:
\[
M_t = \frac{B_t + A_t}{2}
\]

And the next mid-price at time \(t+1\), i.e \textit{NextMidPrice} is given by:
\[
M_{t+1} = \frac{B_{t+1} + A_{t+1}}{2}
\]

Finally, we want to generate a quote price that is within \$0.01 range of the market mid-price \( P_{mid} \) and followed these steps:
\begin{enumerate}
    \item Let \( P_{mid} \) be the market mid price.
    \item Generate a random number \( \delta \) from a uniform distribution within the interval \([-0.01, 0.01]\).
    \item The quoted price \( P_{quote} \) is then given by:
    \[
    P_{quote} = P_{mid} + \delta
    \]
\end{enumerate} 

\section{Data Processing and Feature Engineering}
\label{Data Processing and Feature Engineering}
The target variable we want to predict is \textit{Status}. We constructed a feature set that excludes the unknown variable \textit{NextMidPrice}, the target \textit{Status} and we also created several new features based on exiting variables.

The momentum factor in financial markets refers to the tendency of securities to continue in their existing trend for a period of time. For intraday data, momentum can be defined by comparing the current price to a previous price within the same trading day. If \( P_t \) represents the price at time \( t \) and \( P_{t-k} \) represents the price \( k \) minutes earlier, the momentum factor \( M_t \) can be defined as:
\[
M_t = \frac{P_t - P_{t-k}}{P_{t-k}}
\]
This formula measures the relative change in price over the intraday period from \( t-k \) to \( t \).

We computed momentum factor lagging 5, 10 and 20 periods as the feature: \textit{MOM5}, \textit{MOM10}, \textit{MOM20}. We also calculated \text{Spread} as the difference between \text{MidPrice} - \text{QuotedPrice}. It reflects the relative bid-ask pressure in the market. Spreads can widen during periods of high market volatility. Finally, we take the $log$ term over bond notional to create \text{LogNotional}.

\[
\begin{aligned}
    \text{MOM5} &= \frac{\text{MidPrice}(i)}{\text{MidPrice}(i-5)} - 1 \\
    \text{MOM10} &= \frac{\text{MidPrice}(i)}{\text{MidPrice}(i-10)} - 1 \\
    \text{MOM20} &= \frac{\text{MidPrice}(i)}{\text{MidPrice}(i-20)} - 1 \\
    \text{Spread} &= \text{MidPrice} - \text{QuotedPrice} \\
    \text{Response} &= \text{Spread} \times [1 \text{ if bid else } -1] \\
    \text{LogNotional} &= \log(\text{Notional})
\end{aligned}
\]
where \(\text{MidPrice}(i)\) is the \(i\)-th (current) mid-price of a specific bond. We took logarithm transformation on the Notional.

We intend to keep features intuitive and explainable. A relatively small set of classic academic features can help us better understand and  differentiate our model performance and their property in explainability.

\begin{figure}[h]
    \centering
    \includegraphics[width=240pt]{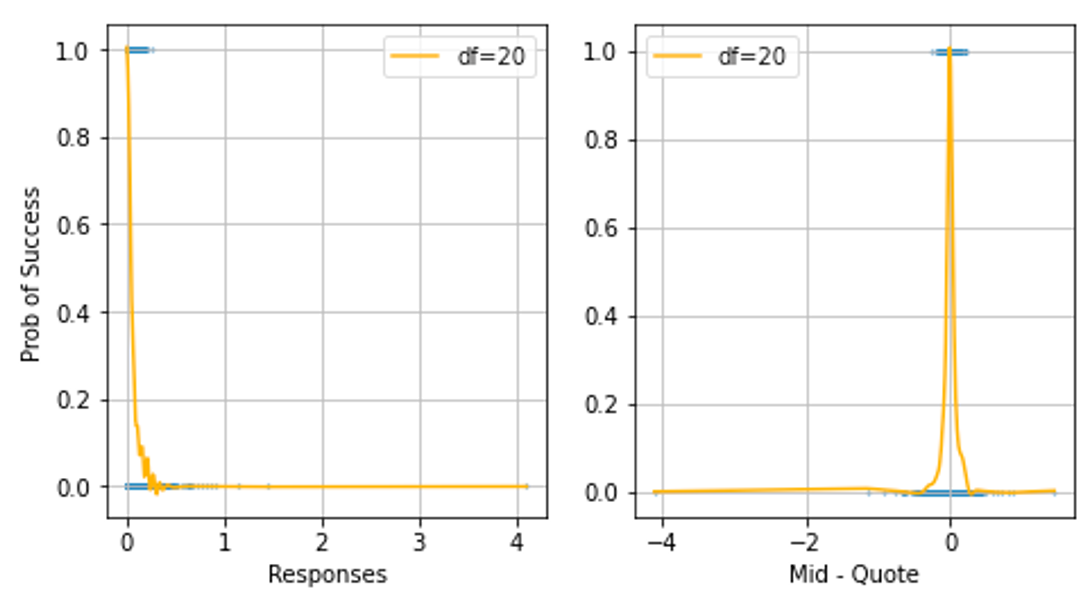} 
    \caption{Kernel probability distributions for each feature}
    \label{fig:splineregression1}
\end{figure}

\begin{figure}[h]
    \centering
    \includegraphics[width=240pt]{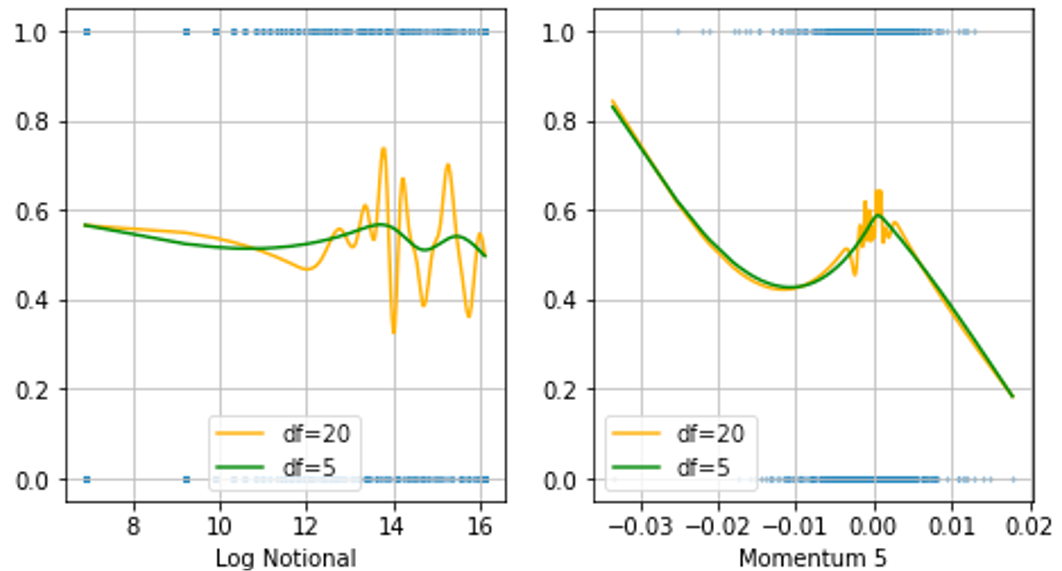} 
    \caption{Kernel probability distributions for each feature}
    \label{fig:splineregression2}
\end{figure}

We used spline regression to calibrate the kernel probability distributions of each feature shown in figure \ref{fig:splineregression1} and figure \ref{fig:splineregression2}. The calibrated distribution confirmed our hypothesis that the probability of Getting Filled tend to decrease when \textit{responses} or \textit{logNotional} increases. We found that \textit{MOM5} is highly correlated with the probability of Getting Filled, but the correlation is not linear and we need to take \textit{Response} into consideration.

\section{Probability of Execution a Trade For a Given Price}
\subsection{Model and Algorithm}
\label{Model}
We first recognize that predicting the probability of winning an RFQ serves as a binary classification problem. Therefore we implemented the following models: Lasso Logistic, Random Forest, XGBoost and Bayesian Neural Network. In addition, we applied 3 groups of ensemble learning: each model predicts the probability of getting RFQ filled and we performed majority vote to generate the final prediction of the probability. Ensembling successfully mitigates the problem overfitting and improve accuracy on out-of-sample data. 

We evaluated the performance each model by their ability to minimize the log-loss (cross-entropy). The negative log-loss for a binary outcome \( y \) is defined as:
\[
- (y \log(p) + (1 - y) \log(1 - p))
\]
where \( p \) is the model’s predicted probability of \( y \) belonging to class 1.

The rationale of our selection of models are as follow: we first use Lasso Logistics model to establish a baseline. Lasso can automatically select meaningful features and provides clear explanation of the influence of each variable to the final outcome. However, Lasso may suffer from under-fitting as it is a linear model. 

Secondly, we implemented both bagging-based Random Forest model and boosting-based XGBoost model. These tree models are non-linear in nature and can generally help boost the prediction accuracy compared to a linear model. However they both suffer from the trade-off between performance and explainability. Random Forest is an ensemble of many decision trees. While a single decision tree can be easily visualized and understood, combining many trees makes the overall model more complex and harder to interpret. XGBoost builds trees sequentially where each new tree aims to correct the residuals of the previous trees. This iterative process makes it hard to interpret the contribution of each feature to the final prediction. 

Finally, we input explore the Bayesian Neural Network model. There are two reasons for the choice of Bayesian Neural Network. Bayesian Neural Network has a probabilistic construct that can naturally apply to our probability prediction task. The model can produce an explainable ratio of likelihood and furthermore, evaluate each new split as a ratio of likelihoods that can balance the trade-off between statistical improvement and the additional model complexity. Moreover, the model takes advantage of both neural network and decision tree. The model has the explainability of decision tree and also has the plasticity of neural networks that allows the node split to have a soft margins and slanted boundaries which can be adjusted by using gradient descent.

\subsection{Bayesian Neural Tree}
Suppose we are tasked with estimating the probability of an outcome \( y \in \mathbb{R} \) given an input \( x \in \mathbb{R}^d \), denoted as \( p(y \mid x) \). 

We define the Bayesian Neural Tree based on a soft Bayesian Decision Tree in \cite{nuti2019bayesian}, Adaptive Bayesian Reticulum as detailed \cite{nuti2020adaptive} and the Bayesian Inference framework in introduced in \cite{zhou2024application}. We propose to integrate Bayesian Inference and Neural Network optimization techniques into the Decision Tree model.

Decision tree on a binary classification problem is a binary tree that can modeled as directed acyclic graphs (DAG). Decision trees consist of two types of nodes: internal nodes \( n \in \mathcal{N} \) and leaves \( \ell \in \mathcal{L} \). We note the root node as \( n_0 \). The set \( \mathcal{L} \) partitions the feature set \( \mathbb{R}^d \) based on parameters associated with the internal nodes \( n \in \mathcal{N} \). 

Internal nodes create certain rules based on an input feature \( x \)  that generate their child nodes and this process can be represented as gate functions \( g_j \) at node \( n_j \) in the Neural Network, where \( j \) indexes the node. Gate functions will produce binary outputs \( \{0, 1\} \). If \( g_j(x) = 1 \), then the left child of an internal node is activated, otherwise the right child is activated.

We define the following probability:
\begin{itemize}
    \item \( p(x \in \ell) \): The probability that input \( x \) falls into leaf \( \ell \), abbreviated as \( p(X \in \ell \mid X = x) \).
    \item \( p(y \mid x \in \ell) \): The probability of outcome \( y \) given that \( x \) is in leaf \( \ell \), abbreviated as \( p(y \mid X \in \ell; X = x) \).
\end{itemize}

Combining these terms with the total law of probability, we can express \( p(y \mid x) \) as:
\[
p(y \mid x) = \sum_{\ell \in \mathcal{L}} p(y \mid x \in \ell) \cdot p(x \in \ell).
\]

In a decision tree, \( p(x \in \ell) \) is calculated as the product of the gate functions along the path from the root node \( n_0 \) to leaf \( \ell \). 

For example:
\[
p(x \in \ell_1) = g_0(x) \cdot g_1(x)\]
\[
p(x \in \ell_2) = g_0(x) \cdot (1 - g_1(x)),
\]
\[
p(x \in \ell_3) = (1 - g_0(x)) \cdot g_2(x)\]
\[p(x \in \ell_4) = (1 - g_0(x)) \cdot (1 - g_2(x)).
\]

\( p(x \in \ell) \) falls between 0 and 1. The probability \( p(y \mid x \in \ell) \) is specified for each leaf.

Decision trees can be seen as single-layer neural networks that maintain a predefined tree structure constraint. The gate function at node \( n_j \) is defined by a manifold \( M \) such that \( g_j(x) = s(d_M(x)) \). \( d_M(x) \) is the signed distance function from \( x \) to \( M \) and \( s \) is the activation function that maps a real value to the range \([0, 1]\). For a hyperplane \( M \) with weight vector \( \mathbf{w} \), we will have \( d_M(x) = \mathbf{w}^T \mathbf{x} + w_0 \). Traditional decision tree has hard boundary, meaning that it defines \( s(x) = \mathbf{1}_{x > 0} \) for gate activation. In the Bayesian Neural Tree, a smooth activation function \( s \) will be better as it allows \( p(x \in \ell) \) to be a continuous variable between 0 and 1 as a function of \( \mathbf{w} \), facilitating the parameter optimization through Nueral Network's backpropagation.

In soft decision trees, \( p(x \in \ell) \) represents the probability that the outcome at \( x \) comes from the distribution associated with leaf \( \ell \). For convenience, we keep the notation of \( x \) belonging to \( \ell \).

We then define a Bayesian Neural Network Framework for a binary classification task. Assume we have data samples \(\{x_i\}_{i=1}^m\) in \(\mathbb{R}^d\) with outcomes \(\{y_i\}_{i=1}^m\) in \(\{0, 1\}\). It is impractical to perform an exhaustive search for the optimal dichotomy for separating \(\{x_i\}_{i=1}^m\) due to the exponential growth of \(m\). Therefore, we need to relax the activation function in decision trees to apply a gradient ascent algorithm.

We use affine hyperplanes with a sigmoid function \(s\) and define the gate function as: 
\[
g(x) = \frac{1}{1 + e^{-(w_0 + \sum_{k} w_k x_k)}}
\]

Denote \(|\mathcal{N}|\) as the number of internal nodes and \(|\mathcal{L}|\) as the number of leaves. Each data sample is assigned to a unique leaf \(\ell\) in \(|\mathcal{L}|\) different configurations. Each configuration has an outcome \(\omega\) that maps point \(i\) to leaf \(\omega(i)\), so probability \(p(\omega) = \prod_i p(x_i \in \omega(i))\). Assume that we have beta conjugate prior parameters \(\alpha\) and \(\beta\), the posterior parameters for a leaf \(\ell\) and configuration \(\omega\) are:
\[
\alpha'_{\ell}(\omega) = \alpha + \sum_{i=1}^m \mathbf{1}_{\omega(i) = \ell} (1 - y_i)
\]
\[
\beta'_{\ell}(\omega) = \beta + \sum_{i=1}^m \mathbf{1}_{\omega(i) = \ell} y_i
\]

Therefore the final log-likelihood is \[
\sum_{\omega \in \Omega} \sum_{\ell \in \mathcal{L}} \ln \left( B(\alpha'_{\ell}(\omega); \beta'_{\ell}(\omega)) \right) p(\omega) 
\], where\(B\) is the Beta function.

The expected probabilities at each leaf are:
\[
p(y = 0 \mid x \in \ell) = \frac{\alpha'_{\ell}}{\alpha'_{\ell} + \beta'_{\ell}}
\]
\[
p(y = 1 \mid x \in \ell) = \frac{\beta'_{\ell}}{\alpha'_{\ell} + \beta'_{\ell}}
\]

Note that these probabilities are determined by \(p(x_i \in \ell)\) and \(y_i\), unlike standard neural networks where the final layer is parameterized and optimized through gradient descent. 

To find the optimal Bayesian Neural Network, we need to maximize the log-likelihood. We applies the Jensen Inequality to achieve a lower bound to be optimized:

\[c = \sum_{\ell \in \mathcal{L}} \ln \left( B(\alpha'_{\ell}; \beta'_{\ell}) / B(\alpha; \beta) \right)
\]
with
\[
\alpha'_{\ell} = \alpha + \sum_{i=1}^m p(x_i \in \ell)(1 - y_i)
\]
\[
\beta'_{\ell} = \beta + \sum_{i=1}^m p(x_i \in \ell) y_i
\]

The probabilities at each leaf can also be approximated as:
\[
p(y = 0 \mid x \in \ell) = \frac{\alpha'_{\ell}}{\alpha'_{\ell} + \beta'_{\ell}}
\]
\[
p(y = 1 \mid x \in \ell) = \frac{\beta'_{\ell}}{\alpha'_{\ell} + \beta'_{\ell}}
\]

Finally, we define the unexplained potential for each leaf \(\ell\) as \(c_{\ell} = \ln \left( \frac{B(\alpha'_{\ell}; \beta'_{\ell})}{B(\alpha; \beta)} \right)\). This value serves as a lower bound on the log-probability of the data at leaf \(\ell\) and helps determine which leaf to expand further. A smaller \(c_{\ell}\) indicates a higher potential for improving the log-likelihood.

When maximizing the bound \( c \) defined above, we want to adaptively built from the data instead of predefining a certian structure using the following steps:

We first begin with a vanilla Bayesian Neural Network structure starting at the root node. As the network grows, we randomly select a leaf to extend based on its unexplained potential \( c_{\ell} \). The selection is made with probabilities proportional to \( c_{\ell} \), given by:
    \[
    \text{Probability} = \frac{c_{\ell}}{\sum_{\ell^* \in \mathcal{L}} c_{\ell^*}}
    \]

Then for the selected leaf, we sample some new weights uniformly and optimize them locally for this leaf while keeping others constant. We subsequently perform a global optimization to fine-tune the entire model. We also adopts the network post-pruning. After global optimization, we evaluate each leaf to decide if it should be pruned. Retain leaves that enhance the model's log-likelihood compared to their parent leaf.

There are several key hyperparameters that will effect the convergence of gradient descent:
    \begin{itemize}
        \item \textbf{Stiff Nodes:} Nodes that behave like Heaviside functions, resulting in minimal gradient changes.
        \item \textbf{Soft Nodes:} Nodes with gradients that are more sensitive and need to become stiffer to improve the unexplained potential \( c_{\ell} \).
    \end{itemize}

In summary, The entire training process involves two phases: local optimization focuses on updating the weights of the current leaf while global optimization adjusts all nodes to enhance the overall model. Our Algorithm \ref{algo} integrates local adjustments typical of decision trees with global optimization methods from neural networks.

\begin{algorithm}[H]
\caption{Construction of Bayesian Neural Network}
\label{algo}
\begin{algorithmic}[1]
\STATE \quad $a = 0$, initialize the vanilla network structure
\WHILE{$a < \text{max\_attempts}$}
    \STATE \quad $j \gets$ choose the index of node to extend
    \STATE \quad Sample the weights for node $j$
    \STATE \quad Train node $j$'s weights using gradient ascent
    \STATE \quad Train the full tree’s weights using gradient ascent
    \STATE \quad Prune the network
    \STATE \quad $a = a + 1$
\ENDWHILE
\end{algorithmic}
\end{algorithm}

\subsection{Choice of Model Parameter}
To optimize the model performance, we first predetermine a range for each parameter based on financial rationale and rule of thumb in tuning. Then we perform a 5-fold time series cross-validation and optimize parameters with an exhaustive grid search. All models are trained in a rolling window fashion on the the training dataset and then fine-tuned on the validation dataset. 

We specially discuss the details of choosing parameter ranges for the Bayesian Neural Tree.
\begin{enumerate}
    \item \textbf{Prior:} For the binary classification, we can use prior distribution of Beta distribution with an $(\alpha, \beta)$ prior. We select Beta(1,1) as the prior to assume that the probability is around 50\%, but with little confidence.
    
    \item \textbf{pruning\_factor:} It is the factor by which the log likelihood of a split has to be larger than the log likelihood of no split in order to keep the split during the pruning stage. We choose between [1,1.05] as the pruning factor.
    
    \item \textbf{n\_iter:} It is the number of attempts to add new nodes by splitting existing ones. The resulting tree will at most contain n\_iter nodes, but it may contain less due to pruning. Larger values lead to more complicated models at the cost of potentially overfitting and slower training performance. We choose between [50,500].
    
    \item \textbf{learning\_rate\_init:} The initial gradient descent learning rate of Adam optimizer and we set it to be 0.05 as a rule of thumb.
    
    \item \textbf{n\_gradient\_descent\_steps:} The number of gradient descent steps to perform in each iteration. Half of the steps will perform local gradient descent on only the newly added node and half will be applied during global gradient descent involving the whole tree. We choose the value of 100.
    
    \item \textbf{initial\_relative\_stiffness:} Small values such as from 0.1 to 5 represent "soft" splits allowing for further optimization through gradient descent. Large values (above 20) make the split an almost step function and disable the weight update. We tested both small and large stiffness values and decide to choose a small stiffness between 2 to 10 to give a soft split that allows for further optimization through gradient descent.
\end{enumerate}

\subsection{Out-of-sample Test and Model Comparison}
\label{Model Comparison}
\begin{figure}[h]
    \centering
    \includegraphics[width=150pt]{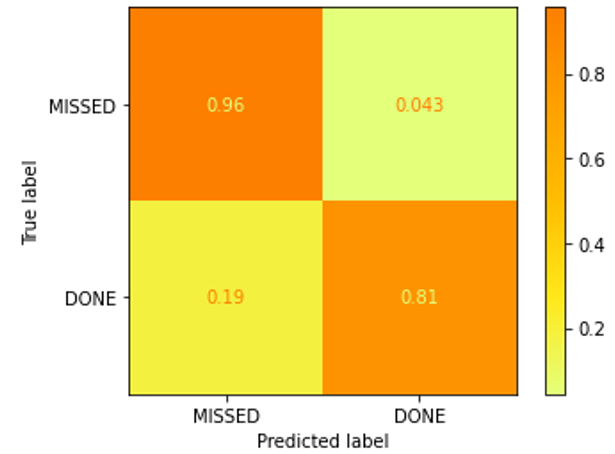} 
    \caption{Confusion Matrix of Bayesian Neural Network}
    \label{fig:confusion}
\end{figure}

\begin{figure}[h]
    \centering
    \includegraphics[width=150pt]{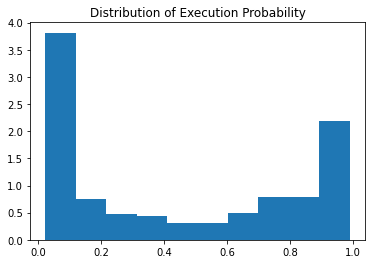} 
    \caption{Distribution of Prediction of Bayesian Neural Network}
    \label{fig:distribution}
\end{figure}

\begin{figure}[h]
    \centering
    \includegraphics[width=220pt]{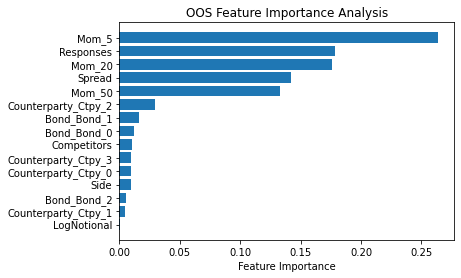} 
    \caption{Feature Importance of Bayesian Neural Network}
    \label{fig:importance}
\end{figure}

\begin{figure}[h]
    \centering
    \includegraphics[width=200pt]{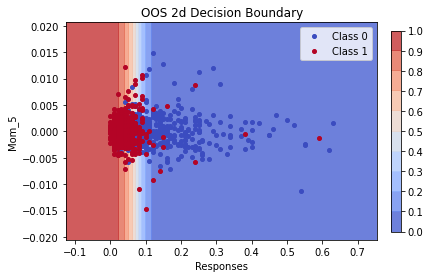} 
    \caption{Decision Boundary of Bayesian Neural Network}
    \label{fig:boundary}
\end{figure}

We split the 10000 samples into 70\% training and 30\% test data. The Bayesian Neural Network model accuracy on out of sample dataset is around 90\%. As shown in figure \ref{fig:confusion}, the False Positive Rate (FPR) is smaller than False Negative Rate, which means that the percentage of RFQs predicted as filled (done) but is actually missed is higher. Further, we found that there are more errors occur when we have more than 2 competitors. The forecasted execution probability is well concentrated to 0 and 1 according to figure \ref{fig:distribution}. As we can see in figure \ref{fig:importance}, the most importance features are \textit{response}, \textit{MOM5}, \textit{MOM20} and \textit{counterparty}. To analyze all possible splits within a dimension, i.e. produce a decision boundary, all we need to do is sort the data once for each dimension. According to the decision boundary in figure \ref{fig:boundary},
the \textit{response} feature is an important factor in separating the filled RFQ from unfilled ones.

Then we compared the performance of all 9 models: Logistic Regression with Lasso regularization (LR), Bayesian Neural Tree with relative stiffness = 2 (ABR2) and relative stiffness = 6 (ABR6), XGBoost (XGB), Random Forest (RF), Ensemble of LR, ABR2, ABR6 (Ensemble 1), Ensemble of LR, ABR6, XGB (Ensemble 2) and Ensemble of LR, ABR2, ABR6, RF, XGB (Ensemble 3). We finally choose the model Ensemble 1 as it has the highest accuracy and F1 score: 

\begin{figure}[h]
    \centering
    \includegraphics[width=240pt]{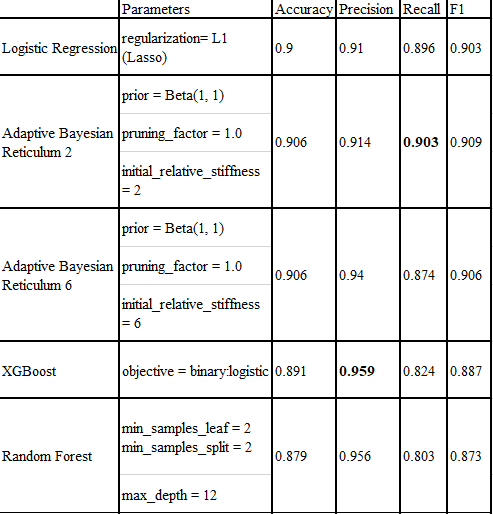} 
    \caption{Model Parameter and Performance}
    \label{tab:compare1}
\end{figure}

\begin{figure}[h]
    \centering
    \includegraphics[width=240pt]{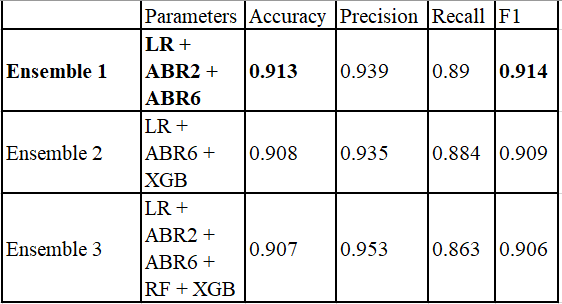} 
    \caption{Model Parameter and Performance}
    \label{tab:compare2}
\end{figure}

\section {Provide a Quote For a Hypothetical RFQ}

\subsection{Next Mid-Price Prediction}
\label{price prediction}
In the intraday trading, \textit{NextMidPrice} is highly correlated with the current mid price \textit{MidPrice}, with a difference followinglog-normal distribution. Assume that the data generation process is unknown, we can still confirm this fact using Q-Q plot as the shown in figure \ref{fig:QQplot}. Therefore, we fitted a linear regression model on \textit{MidPrice} and \textit{Side} to predict \textit{NextMidPrice} on the training set and the adjusted R-squared is 98\% on the validation set. 

\subsection{Market-Maker Utility}
We define the following rules to simulate a real-world RFQ enviroment:

1. Each RFQ that would incur a loss (i.e., an exccessively competitive price) will generate reduce utility by 1 point and be removed from the list of eligible RFQ responses:
\[
\text{Utility} = -1 \quad \text{if loss}
\]

2. Out of all the remaining eligible RFQ responses (i.e., the ones that would have a positive, or at least zero, profit if executed), only the most competitive price wins the RFQ and adds 1 to the utility:
\[
\text{Utility} = +1 \quad \text{for the most competitive price}
\]

3. If multiple RFQs are filled at the price at the exact same level, these market-makers will share the equally 1 utility point but each minus an anti-competition penalty of 0.5:
\[
\text{Utility} = \frac{1}{n} - 0.5 \quad \text{if } n \text{ responses have the same price}
\]
where \( n \) is the number of RFQs priced at the same level.

\bigskip

Putting it all together, we get:
\[ 
\text{Utility} = 
\begin{cases} 
-1 & \text{if loss} \\
+1 & \text{the most competitive price} \\
\frac{1}{n} - 0.5 & \text{if } n \ \text{RFQs have the same price}
\end{cases}
\]
Based on the rules, our goal is to find the best price to fill an RFQ while maximizing the expected market-maker's utility. We convert this problem to an optimization problem by defining the objective function as the expected pay-off:

\begin{figure}[h]
    \centering
    \includegraphics[width=270pt]{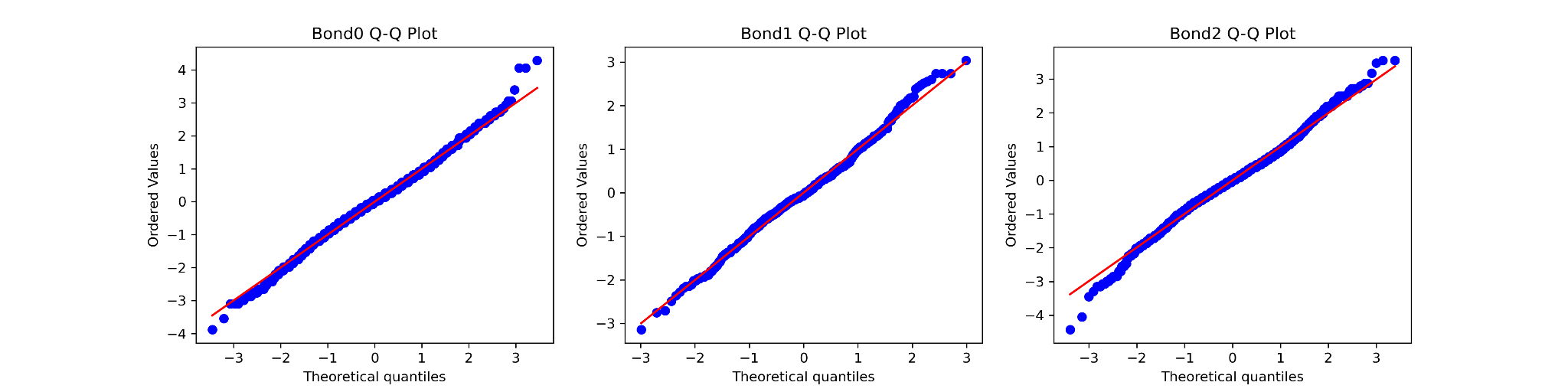} 
    \caption{Correlation of Mid-price and Next Mid-price}
    \label{fig:QQplot}
\end{figure}

\[E(\text{payoff}) = P(\text{Getting Filled})
+ (-1) \times P(\text{Exceed Limit})\]

Note that our expected payoff objective does not take into account the penalty term when other market-makers produce an RFQ that has the exact same quote price with us, as it involves constructing other complicated multi-agent models and is out of the scope of this paper. As an alternative, we assume that probability of having an RFQ of the same price level is negligible and we use a qualitative approach to adapt to the penalty..

\subsection{Out-of-Sample Performance}
To optimize the objective function, we can use the Ensemble 1 model in Section \ref{Model Comparison} to predict the probability of getting the RFQ filled: \( P(\text{Getting Filled}) \). We then use the sample mean of \textit{NextMidPrice} on the validation data to estimate the probability of incurring a loss: \( P(\text{Exceed Limit}) \) for each TBA bond given a quote price. 

Take TBA Bond $0$ as an example, after we predict the \textit{NextMidPrice} \( \hat{P^i} \) of an RFQ \( i \) on the validation set using the linear model in Section \ref{price prediction}, we start a search for the optimal quote price in an interval: \( [\hat{P^i} - 1, \hat{P^i}- 0.99, \ldots, \hat{P^i} + 0.99, \hat{P^i} + 1] \) with an increment of 100bps. As a market-maker, we should have the quote price below \( \hat{P^i} \) for a bid RFQ, and for an ask RFQ, the quote price should be above \( \hat{P^i}\). If our quote price exceeds the true \textit{NextMidPrice} \( P^i\), we have 
\( I(\text{Exceed Limit}) = 1 \); otherwise, \( I(\text{Exceed Limit}) = 0 \) in the expected payoff function.

\begin{figure}[h]
    \centering
    \includegraphics[width=240pt]{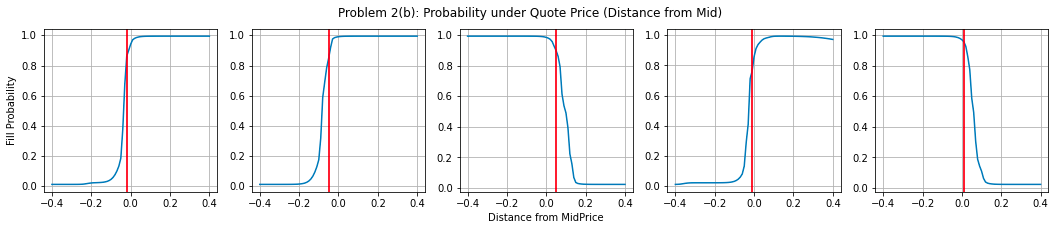} 
    \caption{Probability of Filling an RFQ Given Quote Price}
    \label{fig:probability}
\end{figure}

The probability of incurring a loss given a quote price $P$: \( P(\text{Exceed Limit} \mid P) = \text{Avg}_i [I(\text{Exceed Limit} \mid P^i)] \) for all sample RFQ \( i \) in the validation dataset. For each TBA bond, we can generate a series of pairs of this possibility and our quote price. We show the \( P(\text{Exceed Limit}) \)-to-price curve on validation data in figure \ref{fig:probability} and use it for the test set. The red lines represent our final decision of quote price. According to the plot, our RFQ with the selected quote price have a high probability of getting filled according to the Ensemble 1 model we built in Section \ref{Model Comparison}.

\begin{figure}[h]
    \centering
    \includegraphics[width=240pt]{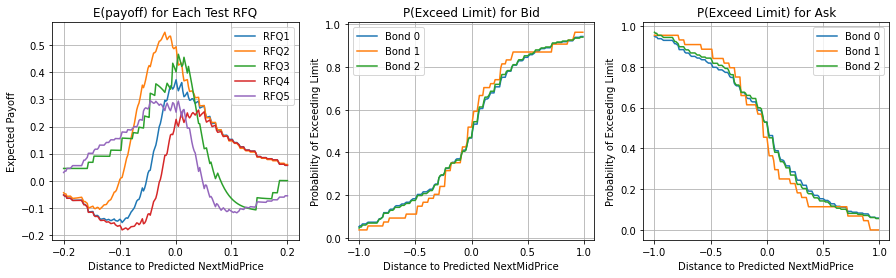} 
    \caption{Market-Maker Utility given Quote Price}
    \label{fig:final prediction}
\end{figure}

To adjust for the penalty of having RFQ at the same price level, we define a cap for both bid and ask RFQ: the final quote price for a bid RFQ is $max(\textit{BidPrice},\textit{MidPrice}-0.01)$ and the final quote price for an ask RFQ is $min(\textit{AskPrice}, \textit{MidPrice}+0.01)$. Our final quote price are: [-0.02, -0.05, 0.05, -0.01, 0.01] for 5 out-of-sample RFQs. As shown in figure \ref{fig:final prediction}, we define the distance to \textit{NextMidPrice} as our quote price minus the prediction of next Mid-price. The expected payoff (utility) for market-maker peaks at a small interval around the true \textit{NextMidPrice}. The probability of filling a bid RFQ is at the highest when we set the quote price far above the true \textit{NextMidPrice} (it means our quote price is highly competitive at the price of incurring a loss for market makers). Similary, the probability of filling an ask RFQ is at the highest when we set the quote price far below the true \textit{NextMidPrice}.

We simulate the quote price of other market makers using the same data generation process in Section \ref{Data Generation Process}. Among all 5 RFQs out of sample, we are able to fill 4 RFQ and secure positive utility values on all 5 of them. It shows that our prediction of RFQ fill rate and mid-price level using XAI is useful for market makers.

\section{Conclusion}

In this paper, we explore the formulation of RFQ for TBA bond. We first establish a simulation algorithm to genereate the RFQ data process. We then model the probability of getting RFQ filled with multiple machine learning algorithms and explainable AI models including LASSO Logistic Regression, XGBoost and Bayesian Neural Tree. We found that Bayesian Neural Tree has balanced off a good performance and a strong explainability with its hierarchical tree structure. An ensemble of these models further boost its prediction accuracy. Finally, we propose a market-maker's utility function to optimize the best quote price. The paper makes practical implications on providing a framework for analyzing RFQ data and promoting pricing efficiency for TBAs through RFQ to boost the mortgage lending process.


\end{document}